# Correlating Quasi-Optical Coupling Efficiency with Measured Receiver Noise Temperature in Metalens Coupled THz HEB Mixer

Jingqi Hu, Rui Wang, Jiameng Wang, Kangmin Zhou, Hua Li, and Dingding Ren

***Abstract*—Quasi-optical coupling serves as the critical interface in terahertz (THz) heterodyne receiver systems, enabling efficient transfer of incident radiation to photomixers through a focusing element and a planar microwave antenna. With recent advances in nanofabrication, planar dielectric metalenses have emerged as promising alternatives to conventional refractive optics due to their compactness and scalability. However, unlike conventional elliptical silicon lenses that are often treated as nearly ideal optical components, the focusing characteristics of metalenses, including both phase and amplitude, strongly depend on the local deflection angle across the aperture, creating an urgent need to quantitatively understand the coupling between a dielectric metalens and a planar antenna. In this work, we present a quasi-optical coupling analysis between a planar Si metalens and a logarithmic spiral antenna integrated with a THz superconducting NbN hot-electron bolometer (HEB) mixer operating at 1.63 THz using spherical-coordinate vectorial integration. By combining the angular radiation profile of the spiral antenna with the propagated complex electric-field profile from metalens numerical simulations, the calculated coupling efficiency accounts for angular power distribution, phase-front matching, and polarization-dependent vectorial overlap. The calculated coupling efficiency is then directly correlated with experimentally measured double-sideband receiver noise temperatures through comparison with a conventional elliptical Si lens measured under the same receiver configuration. The analysis establishes a quantitative relationship between metalens focusing efficiency, vectorial antenna coupling, and receiver noise temperature, providing guidance for optimizing metalens design and improving the overall performance of metalens-integrated THz heterodyne receivers.**

## I. INTRODUCTION

Quasi-optical lens–antenna and waveguide-based architectures represent two of the most widely used approaches for coupling electromagnetic fields into superconducting mixers in millimeter- and submillimeter-wave heterodyne receiver systems [1], [2]. In waveguide-based configurations, radiation is first converted into a well-defined guided mode by a feed horn and metallic waveguide, and the mixer samples this mode via a probe antenna embedded within the waveguide structure, effectively coupling to the local electric field of a single-mode waveguide [3], [4]. In contrast, quasi-optical systems operate in free space, where dielectric lenses or reflective optics focus the incident radiation directly onto a planar antenna, which samples the spatially distributed electromagnetic field at the focal region [5], [6], [7], [8]. These two approaches therefore differ not only in implementation but also in the underlying field representation: waveguide systems rely on modal confinement and impedance-controlled coupling, whereas lens–antenna systems couple directly to the angular and spatial distribution of propagating fields [9]. While waveguide-coupled receivers provide precise mode control and high efficiency at lower frequencies, their implementation becomes increasingly challenging at terahertz (THz) frequencies due to fabrication tolerances and scaling limitations [4]. Quasi-optical lens–antenna systems, by avoiding guided-wave constraints, offer a more scalable and fabrication-friendly solution for high-frequency operation and planar integration, which has been employed for the Herschel Space Observatory [10] and balloon-borne observatories, such as STO-2 [11], GUSTO [12] and OASIS-B [13].

In microwave and submillimeter-wave receiver systems, quasi-optical coupling is traditionally analyzed within the framework of Gaussian beam optics [2]. Metrics such as far-field Gaussian beam profiles and Gaussicity are widely employed to characterize beam quality and to guide the optical design,

This work was supported in part by the National Key R&D Program of China (2025YFC2423702), in part by the National Natural Science Foundation of China (62575299, 62325509, 62235019, 62505344, 62531005, 62275258, 62305364, 62435017, and T2550072), and in part by the Quantum Science and Technology-National Science and Technology Major Project (2023ZD0301000) and the Scientific Instrument and Equipment Development Project of the Chinese Academy of Sciences (PTYQ2026YZ0050), Science and Technology Commission of Shanghai Municipality (23ZR1474000), the CAS Project for Young Scientists in Basic Research (YSBR-069), and Autonomous deployment project of the State Key Laboratory of Materials for Integrated Circuits (SKLJC-Z2025-B03).

(*Corresponding author: Dingding Ren*)

Jingqi Hu and Rui Wang contributed equally to this work and are with the State Key Laboratory of Materials for Integrated Circuits and the Key Laboratory of Terahertz Solid-State Technology, Shanghai Institute of Microsystem and Information Technology, Chinese Academy of Sciences, Shanghai 200050, China. (email: jingqi.hu@mail.sim.ac.cn; rui.wang@mail.sim.ac.cn)

Jiameng Wang and Kangmin Zhou are with the Purple Mountain Observatory, Chinese Academy of Sciences, Nanjing 210023, China. (email: jmwang@pmo.ac.cn; kmzhou@pmo.ac.cn)

Hua Li is with the State Key Laboratory of Materials for Integrated Circuits and the Key Laboratory of Terahertz Solid-State Technology, Shanghai Institute of Microsystem and Information Technology, Chinese Academy of Sciences, Shanghai 200050, China and with the Center of Materials Science and Optoelectronics Engineering, University of Chinese Academy of Sciences, Beijing 100049, China. (email: hua.li@mail.sim.ac.cn)

Dingding Ren is with the State Key Laboratory of Materials for Integrated Circuits and the Key Laboratory of Terahertz Solid-State Technology, Shanghai Institute of Microsystem and Information Technology, Chinese Academy of Sciences, Shanghai 200050, China. (e-mail: dren@mail.sim.ac.cn)

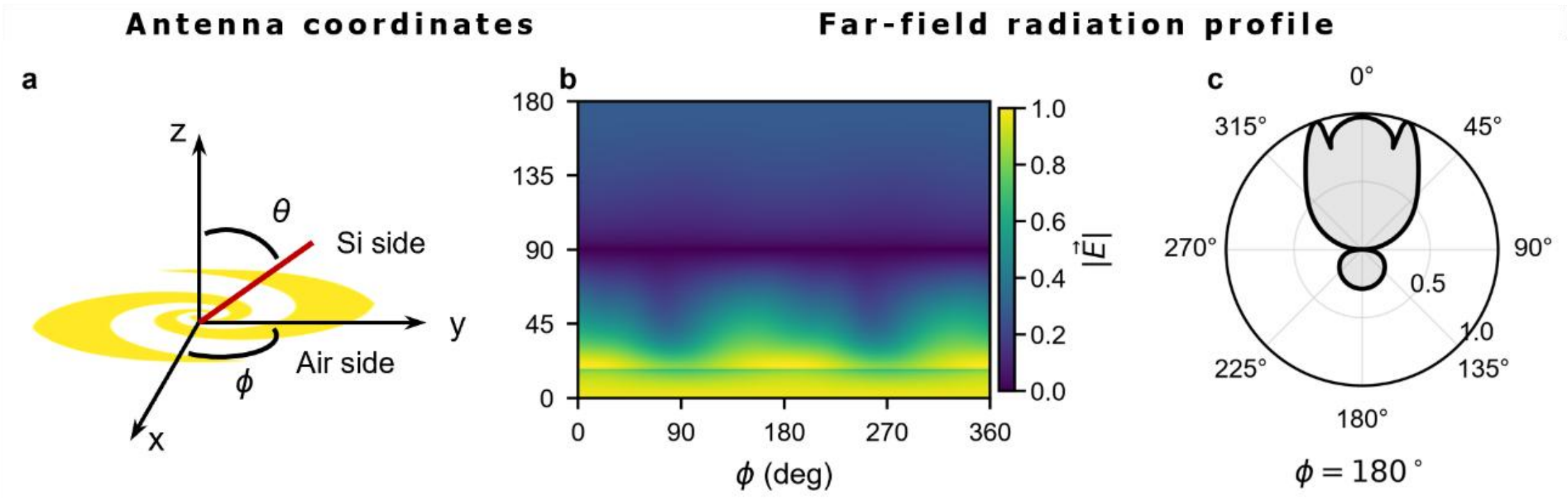

**Fig. 1.** Definition of the antenna coordinate system and corresponding far-field radiation characteristics. **a.** Spherical coordinate system ($\theta$, $\phi$) defined at the Si–air interface, with $\theta$ measured from the surface normal toward the silicon side. **b.** Angular radiation amplitude distribution of the spiral antenna expressed in spherical coordinates. The antenna radiation $\left|\vec{E}_{ant}(\theta, \phi)\right|$ is plotted as a function of polar angle $\theta$ and azimuthal angle $\varphi$, showing a predominantly azimuthally uniform distribution with strongest emission at small $\theta$ (toward the silicon side) and a gradual decay toward larger angles. **c.** Polar profile of the normalized antenna electric-field amplitude $\left|\vec{E}_{ant}(\theta)\right|$ at $\phi = 180°$, plotted on a linear scale.

enabling efficient matching between the receiver beam and the incoming radiation from the telescope [5], [14]. However, these Gaussian-based descriptions are inherently approximate and become increasingly inadequate for non-paraxial or high-numerical-aperture (NA) systems. In particular, establishing a consistent link between plane-wave illumination, used in hot/cold load calibration, and the resulting receiver noise temperature is nontrivial, often requiring elaborate analytical treatments [15], or relying on full-wave electromagnetic simulations for accurate evaluation [7].

This quasi-optical interface becomes further complicated with the emergence of planarized metalens platforms, where both the transmission amplitude and phase response are inherently dependent on the deflection angle across the aperture. In such systems, the conventional assumption of negligible refraction loss is no longer valid, particularly at large deflection angles [16], [17]. As a result, simplified Gaussian-based coupling models become inadequate, and a more rigorous yet tractable framework is required to accurately describe the coupling between the metalens and the antenna.

Recently, we demonstrated the first planarized metalens-coupled THz superconducting hot-electron bolometer (HEB) mixer with a logarithmic spiral antenna operating at 1.63 THz, achieving a double-sideband receiver noise temperature ($T_{rec}^{DSB}$) of 1800 K [18]. By benchmarking against the conventional elliptical Si lens within the same experimental configuration, this performance corresponds to an effective coupling efficiency of approximately 27% relative to the elliptical-lens counterparts. Despite this experimental evidence, a quantitative analysis for the coupling efficiency between the dielectric metalens and spiral antennas remains lacking.

In this work, we employ spherical-coordinate integration to quantitatively evaluate the coupling efficiency between a dielectric metalens and a logarithmic spiral antenna. The analysis uses the propagated complex electric field from metalens simulations and the complex vector far-field distribution of the spiral antenna, thereby accounting for angular field matching, phase-front alignment, and vectorial polarization projection. Because the same HEB mixer and receiver setup are used for both the metalens and elliptical-lens configurations, the observed difference in receiver noise temperature can be attributed primarily to differences in lens transmission and lens–antenna coupling efficiency [18], [19].

For the metalens configuration, the propagated complex field obtained from the metalens simulation is first used to construct the angular amplitude and phase distribution of the metalens-delivered field. This scalar angular response is assigned to different transverse polarization basis vectors and projected onto the local spherical components, $E_\theta$ and $E_\phi$. Different incident linear-polarization states are then represented by coherent linear combinations of these projected basis fields. This treatment accounts for the vectorial projection of the metalens-delivered field onto the antenna receiving mode, while assuming that the scalar transmission and phase response of the metalens are the same for orthogonal incident polarizations due to the limitation of two-dimensional (2D) finite-difference time-domain (FDTD) simulation. The resulting complex spherical field components are combined with the corresponding $E_\theta$ and $E_\phi$ components of the spiral-antenna far-field receiving mode to evaluate the vectorial overlap. This procedure accounts for angular field distribution, phase-front matching, and polarization-dependent mode overlap between the metalens-delivered field and the antenna receiving mode.

Since the hot/cold-load radiation used in the receiver noise-temperature measurement is thermal and unpolarized, the blackbody-relevant coupling efficiency is obtained by averaging the coupling efficiencies of orthogonal incident linear-polarization states. For the conventional elliptical lens, the angular transmission and beam projection are modeled using geometrical optics, incorporating Snell's law and Fresnel coefficients to account for refraction and interfacial transmission. In both cases, plane-wave excitation is used to represent the incident electromagnetic field. By combining the calculated unpolarized vectorial coupling efficiency with the lens transmission characteristics, the overall optical efficiency can be directly correlated with the measured $T_{\mathrm{rec}}^{\mathrm{DSB}}$. This analysis enables a direct comparison between metalens- and elliptical-lens-based coupling, linking angular field distribution, phase-front matching, polarization-dependent vectorial overlap, and experimentally measured receiver noise temperature.

## II. Coupling Efficiency Calculation For Metalens-Spiral Antenna by Spherical Integration

We first evaluate the metalens-antenna coupling efficiency using a spherical-integration analysis based on the coordinate system shown in Fig. 1a, with the coordinate origin located at the HEB microbridge [5], [15]. This choice provides a direct mapping from the Cartesian field distribution to its angular representation in spherical coordinates, which is the natural basis for describing the quasi-optical coupling between the lens and the antenna.

For each propagation direction $(\theta, \phi)$, the transverse electric field can be decomposed into two orthogonal components along the local spherical basis vectors $\vec{e}_\theta$and $\vec{e}_\phi$,

$$\vec{E}(\theta,\varphi) = E_\theta(\theta,\varphi)\vec{e}_\theta + E_\varphi(\theta,\varphi)\vec{e}_\varphi. \quad (1)$$

Here, $E_\theta$ and $E_\phi$ are complex scalar field components, while $\vec{e}_\theta$and $\vec{e}_\phi$ are orthogonal unit vectors. In the calculation, the spiral antenna is described by its full complex far-field receiving mode, including both $E_\theta$ and $E_\phi$ components. Therefore, the lens–antenna coupling efficiency for a specified coherent incident polarization state can be calculated from the normalized complex vector overlap between the metalens-delivered field and the antenna receiving mode [20],

$$\eta_{coupling,linpol} = \frac{\left|\int_0^{2\pi}\int_0^{\pi} E_{lens,\theta}(\theta,\phi)E^*_{ant,\theta}(\theta,\phi)+E_{lens,\phi}(\theta,\phi)E^*_{ant,\phi}(\theta,\phi)d\Omega\right|^2}{\left[\int_0^{2\pi}\int_0^{\pi}\left|E_{lens,\theta}\right|^2+\left|E_{lens,\phi}\right|^2 d\Omega\right]\left[\int_0^{2\pi}\int_0^{\pi}\left|E_{ant,\theta}\right|^2+\left|E_{ant,\phi}\right|^2 d\Omega\right]} \quad (2)$$

where $d\Omega = sin\theta d\theta d\phi$. In Eq. (2), the complex conjugation of the antenna field accounts for the relative phase between the two fields, while the two transverse components, $E_\theta$ and $E_\phi$, account for polarization matching in the local spherical basis. Thus, this formulation includes both phase and vectorial polarization mismatch that is not captured by a scalar power-overlap model.

For unpolarized incident radiation, the raw unpolarized coupling efficiency is first obtained by averaging the efficiencies calculated for two orthogonal incident polarization states,

$$\eta_{unpol} = \frac{1}{2}\left[\eta_{coupling,linpol}(\alpha) + \eta_{coupling,linpol}(\alpha + 90°)\right]. \quad (3)$$

In the numerical implementation, the unpolarized response is obtained by averaging over orthogonal incident-polarization pairs. For each incident linear-polarization angle α, the coupling efficiencies for α and α+90° are first averaged, and the final raw unpolarized coupling efficiency is obtained by averaging this pair-averaged value over α. This procedure avoids selecting a specific incident polarization orientation relative to the azimuthally varying receiving pattern of the spiral antenna.

Because the blackbody radiation can be decomposed into two incoherent orthogonal polarization modes, whereas the logarithmic spiral antenna couples the HEB microbridge to one circularly polarized receiving mode, the raw unpolarized overlap $\eta_{unpol}$ is referenced to the total power in both polarization modes. To compare with the coupling efficiency of the single antenna receiving mode, we define a single-mode-referenced blackbody coupling efficiency as

$$\eta_{coupling,blk} = 2\eta_{unpol}. \quad (4)$$

This factor of two converts the unpolarized lens–antenna coupling efficiency, which is referenced to the total power carried by two orthogonal polarization states of the blackbody radiation, into a coupling efficiency referenced to the single circularly polarized antenna mode coupled to the NbN mixer.

### *A. Radiative Profile of Spiral Antenna*

We first examine the far-field response of the logarithmic spiral antenna using full-wave FEKO simulations. The spherical coordinate system used for the angular representation is defined in Fig. 1a, where the polar angle $\theta$ is measured from the surface normal toward the silicon substrate side and $\phi$ denotes the azimuthal angle. The simulated far-field electric-field distribution is shown in Fig. 1b as a 2D angular map in $(\theta, \phi)$, with a representative polar cut at $\phi = 180°$ shown in Fig. 1c. The antenna response is predominantly concentrated at small polar angles, corresponding to radiation toward the silicon substrate, and gradually decreases with increasing $\theta$. This complex far-field distribution defines the antenna receiving mode used in the vectorial coupling-efficiency analysis described in the following sections.

### *B. Field Propagation through Metalens and Coupling Efficiency Calculation*

Unlike the elliptical lens, the metalens forms the focus through a diffraction-based wavefront-engineering mechanism, and its transmitted field depends strongly on the local deflection angle [16], [17]. To capture the metalens focusing characteristic with manageable computational cost, we performed a 2D FDTD simulation [21], [22] in which a normally incident plane wave $E_z$ propagates through the metalens into the silicon substrate, as illustrated in Fig. 2a. To approximate the contribution of the circular aperture in the 2D simulation, the incident electric-field amplitude was weighted by $\sqrt{|y|}$, where y is the transverse coordinate measured from the optical axis [18]. The exact metalens geometry is reported in ref. [18]. A field monitor placed inside the silicon substrate records the transmitted complex electric field, preserving both amplitude and phase information. This monitor field therefore provides the near-field representation of the metalens output and serves as the input for subsequent propagation analysis for focal spot identification.

The complex field recorded at the monitor plane was propagated using the angular-spectrum method. In this method,

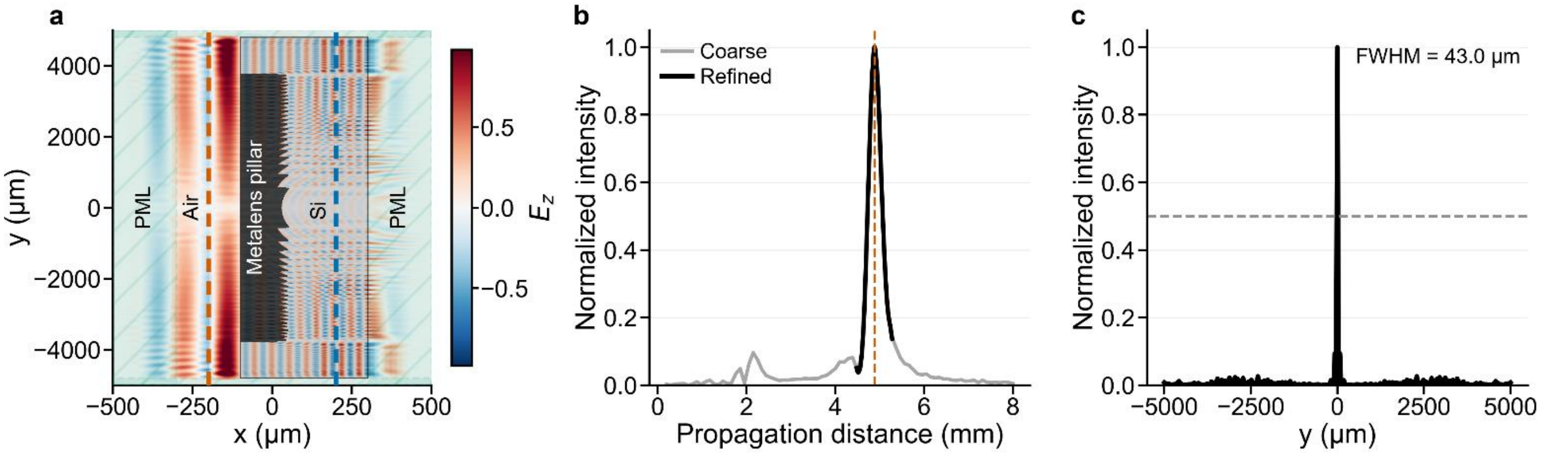

**Fig. 2. Simulated metalens field propagation and focal characterization. a** Instantaneous $E_z$ field distribution in the 2D x-y simulation domain, overlaid with the Si metalens geometry. The black contour outlines the Si/air dielectric boundary, and the region enclosed by the black boundary corresponds to Si. The green dashed/hatched regions indicate the perfectly matched layers (PMLs), while the dashed vertical lines mark the source and monitor planes in red and blue, respectively. **b** Coarse and refined angular-spectrum focus search based on the recorded monitor field, with the extracted focal distance indicated by the dashed vertical line. **c** Normalized focal-plane intensity profile at the extracted focus from **b**. The dashed horizontal line indicates the half-maximum level used to estimate the FWHM.

the monitor-plane field is first decomposed into its angular spectrum $\tilde{E}(k_y)$, and each plane-wave component is then propagated by its longitudinal phase factor [23]. The monitor field can be denoted as

$$E_{z,mon}(y) = |E_{z,mon}(y)|\, e^{i\varphi_{mon}(y)}, \tag{5}$$

where $\varphi_{mon}(y)$ is the phase stored at each transverse position on the monitor plane. After applying the aperture window $W(y)$, the transverse spatial-frequency spectrum can be calculated as

$$\tilde{E}(k_y) = \int E_{z,mon}(y)\, W(y)\, e^{-ik_y \cdot y} dy. \tag{6}$$

Here, $\tilde{E}(k_y)$ represents the complex angular-spectrum amplitude associated with the transverse wave vector $k_y$. Because $E_{z,\mathrm{mon}}(y)$ is a complex field, the amplitude and phase recorded at the monitor plane are both retained in the angular-spectrum propagation. Each transverse wave-vector component $k_y$ corresponds to a plane-wave component inside silicon according to

$$k_y = n_{Si} \cdot k_0 sin\theta, \tag{7}$$

and

$$k_x = [(n_{Si} \cdot k_0)^2 - {k_y}^2]^{\frac{1}{2}}, \tag{8}$$

where $k_0 = \frac{2\pi}{\lambda_0}$. Only propagating components satisfying $|k_y| \le n_{Si}k_0$ are retained. Propagation from the monitor plane $x_{mon}$ to the focal plane $x_{focal}$ was then performed by applying the phase factor $e^{ik_x(x_{focal}-x_{mon})}$. The propagated field at each trial plane was reconstructed as

$$E(x_{focal}, y) = \int \tilde{E}(k_y) e^{ik_x(x_{focal}-x_{mon})} \cdot e^{ik_y \cdot y} dk_y. \tag{9}$$

The focal plane was determined by scanning $x_{focal}$ and identifying the position that maximizes the focused intensity. As shown in Fig. 2b, a coarse search followed by a refined search was used to locate this maximum-intensity position along the optical axis. The transverse intensity profile at the extracted focal plane was then obtained by fixing $x = x_{\mathrm{focal}}$ and evaluating $|E(x_{focal}, y)|^2$ as a function of the transverse coordinate $y$, as shown in Fig. 2c, confirming that the focal position is on the optical axis. Therefore, this extracted metalens focal position ($x_{focal}$, 0) can be used to define the antenna reference position for the subsequent angular coupling analysis.

After identifying the focal spot position, a second FDTD propagation simulation was performed using a uniform incident field to extract the angle-dependent complex response of the simulated 2D metalens cross section. The same angular-spectrum procedure was then applied to the resulting monitor field. Instead of reconstructing the real-space field, the propagated angular spectrum was used to define the complex angular response of the metalens at the referenced focal point,

$$E_{ml}^{ang}(\theta) = \tilde{E}(k_y) e^{ik_x(x_{focal}-x_{mon})} \cdot e^{ik_y \cdot y_{focal}}. \tag{10}$$

Here, $y_{focal}$ is the transverse position of the referenced focal point. The quantity $E_{ml}^{ang}(\theta)$ contains both the angular amplitude and the relative phase of the metalens-delivered field at the metalens focal/antenna reference position. Therefore, the relative phase of the metalens-transmitted field recorded at the monitor plane is preserved in the angular response and directly affects the coherent overlap with the antenna receiving mode

For the spherical-coordinate coupling calculation, the scalar angular response $E_{\mathrm{ml}}^{\mathrm{ang}}(\theta)$ obtained from the 2D metalens simulation was extended to the full spherical angular grid by assuming azimuthal symmetry, such that $E_{\mathrm{ml}}^{\mathrm{ang}}(\theta, \phi) = E_{\mathrm{ml}}^{\mathrm{ang}}(\theta)$. This scalar angular field was then assigned to transverse incident-polarization basis fields and projected onto the local spherical basis to obtain $E_{\mathrm{ml},\theta}(\theta,\phi)$and $E_{\mathrm{ml},\phi}(\theta,\phi)$ for each incident linear-polarization state. These metalens field components were evaluated on the same $(\theta,\phi)$ grid as the simulated antenna receiving-field components, $E_{\mathrm{ant},\theta}(\theta,\phi)$ and $E_{\mathrm{ant},\phi}(\theta,\phi)$, and the metalens–antenna coupling was then calculated using the complex vector overlap in Eq. (2). For a selected incident linear polarization, the local complex vector-field inner product was first evaluated on the full $(\theta,\phi)$ grid as $E_{lens,\theta}(\theta,\phi)E^*_{ant,\theta}(\theta,\phi)+E_{lens,\phi}(\theta,\phi)E^*_{ant,\phi}(\theta,\phi)$ . This quantity represents the local coherent overlap between the metalens-delivered field and the antenna receiving mode. To visualize the angular origin of the coupling, the numerator integrand of Eq. (2) was integrated over $\phi$ at each fixed $\theta$,

$$A(\theta)=\int_0^{2\pi}[E_{lens,\theta}E^*_{ant,\theta}+E_{lens,\phi}E^*_{ant,\phi}]\,sin\theta d\phi. \qquad (11)$$

The total complex overlap in the numerator of Eq. (2) is then obtained by integrating $A(\theta)$ over $\theta$. After global phase alignment, the real part of $A(\theta)$ is plotted in Fig. 3b as the $\theta$-resolved vectorial overlap contribution. Positive values indicate constructive contributions to the coherent overlap, whereas negative values indicate destructive contributions caused by phase or polarization-vector mismatch. For unpolarized illumination, the coupling efficiencies were obtained by averaging orthogonal incident-polarization pairs, and the corresponding $\theta$-resolved contribution is shown in Fig. 3c.

For comparison, a power-only angular overlap was also calculated using only the angular power distributions of the metalens and antenna fields. In this case, the metalens and antenna intensities were defined as

$$S^{ang}_{ml}(\theta,\phi)=|E_{ml,\theta}(\theta,\phi)|^2+|E_{ml,\phi}(\theta,\phi)|^2 \qquad (12)$$

and

$$S^{ang}_{ant}(\theta,\phi)=|E_{ant,\theta}(\theta,\phi)|^2+|E_{ant,\phi}(\theta,\phi)|^2, \qquad (13)$$

respectively. The scalar power-overlap efficiency was then evaluated from the normalized overlap of these two angular power distributions, without including the relative phase or vector-polarization direction. This power-only metric therefore represents an upper-bound estimate for angular power matching.

As shown in Fig. 3, the metalens field and antenna receiving mode overlap well at small polar angles $\theta$. For the selected linear polarization, which maximizes the coupling efficiency, the complex vector-basis overlap gives $2\eta_{\mathrm{lin}}$ as high as 26.3 %, while the unpolarized vectorial coupling obtained by averaging two orthogonal incident polarization states gives $2\eta_{\mathrm{unpol}}=25.9\%$. As $\theta$increases, the metalens contribution becomes less smooth, and amplitude fluctuations appear in the angular coupling curve. More importantly, near $\theta\approx30°$, the signed vectorial contribution begins to change sign at specific angles. Its onset indicates that the high-deflection-angle region of the metalens introduces significant phase error relative to the antenna receiving mode.

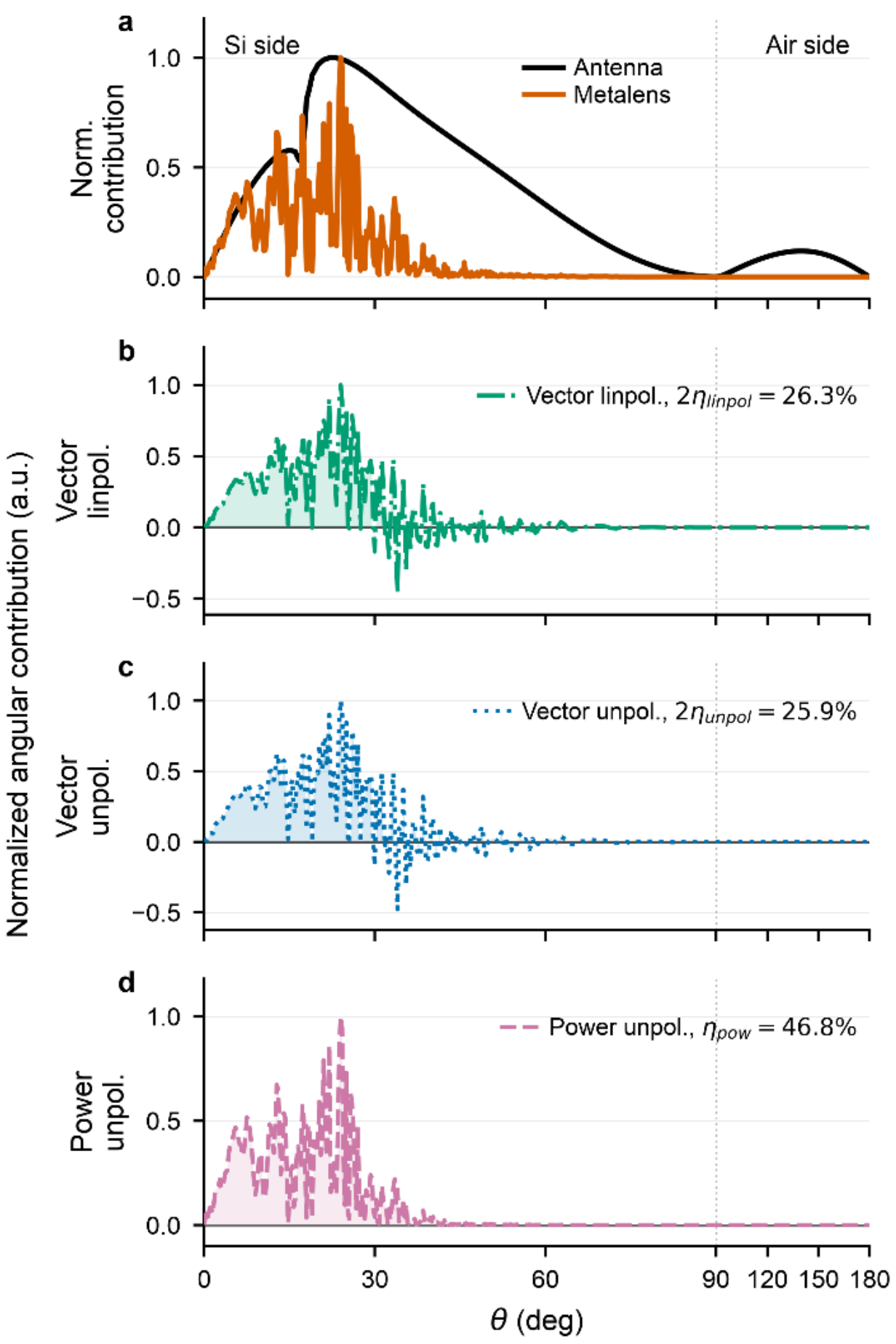

**Fig. 3.** Angular metalens–antenna coupling analysis. **a** Peak-normalized angular power distributions of the spiral antenna and metalens field after azimuthal integration. **b** Coherent vectorial coupling contribution for a selected linear incident polarization. **c** Vectorial coupling contribution for unpolarized illumination, calculated by incoherent averaging over two orthogonal polarization states. **d** Power-only angular overlap between the metalens and antenna power distributions.

This interpretation is supported by the comparison with the power-only overlap in Fig. 3d. The power-only metric gives $\eta_{\mathrm{pow}}=46.8\%$, which is larger than the vectorial unpolarized result because it only compares the angular power-density distributions of the metalens and antenna. It remains positive over the full angular range and therefore cannot capture the destructive contribution caused by phase mismatch or polarization-vector mismatch. The difference between $\eta_{\mathrm{pow}}$ and $2\eta_{\mathrm{unpol}}$ shows that the coupling loss is not only caused by insufficient angular power overlap but is also strongly affected by coherent vectorial mismatch at larger deflection angles, especially near and above $\theta\approx30°$. Our analysis indicates that the reduction in quasi-optical coupling efficiency using metalens is not solely caused by the decrease in metalens focusing field amplitude at high deflection angles. This high-angle region can also contribute destructively to the coherent vectorial overlap because of phase errors and polarization mismatch. This finding is consistent with previous theoretical

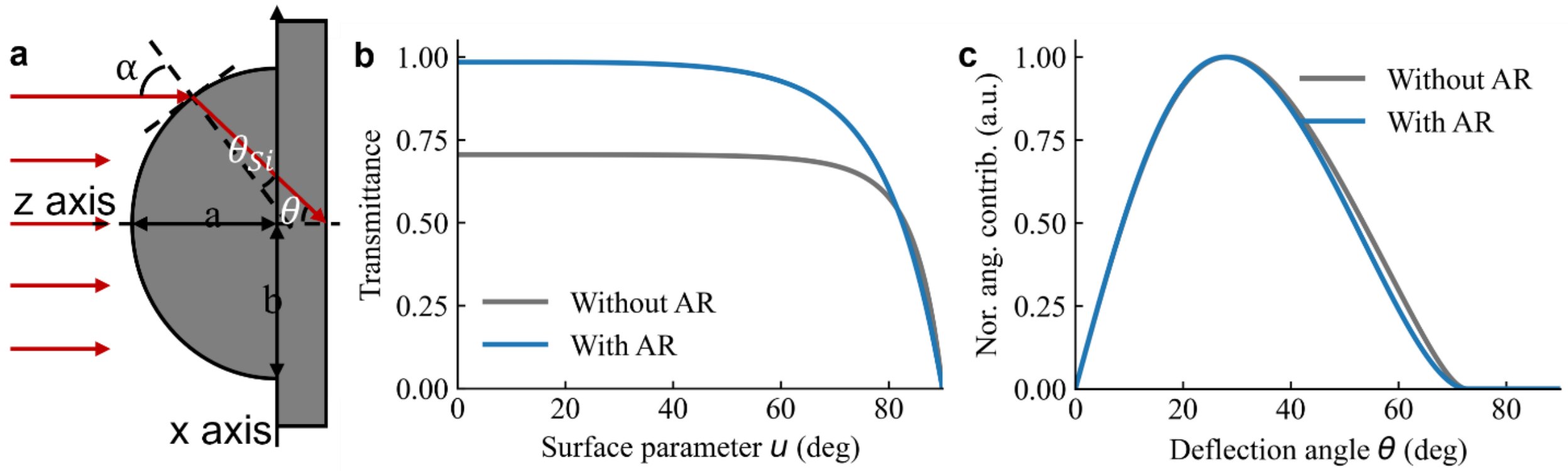

**Fig. 4.** Elliptical lens modeling and angular power-density distribution. **a** Geometrical configuration of the silicon elliptical lens under plane-wave illumination, showing refraction into the Si substrate according to Snell's law and the relevant geometrical parameters. **b** Angle-dependent transmittance of the air–silicon interface as a function of the surface parameter $u$, comparing the cases without anti-reflection (AR) coating and with a Parylene-C AR coating. **c** Corresponding normalized angular power density inside silicon. The AR coating enhances the transmittance over a broad angular range, increasing the total optical power coupled into the Si lens from 72.1% to 92.6%.

studies showing that metalens efficiency deteriorates at large deflection angles [16].

## III. Angular Power-Density Distribution of the Elliptical Lens under Plane-wave Illumination

### *A. Polarization-Averaged Transmittance of the Elliptical Lens*

To validate the coupling-efficiency analysis developed for the metalens, we apply the same angular-overlap formalism to a conventional silicon elliptical lens coupled to the logarithmic spiral antenna. The elliptical lens is modeled using a geometrical-optics-based spherical mapping with the lens dimensions reported in Ref. [19]. As illustrated in Fig. 4a, a plane wave incident from air illuminates the curved lens surface. The optical power intercepted by each surface element is weighted by the projected-area factor cos$\alpha$, where $\alpha$ is the angle between the local surface normal and the incident direction. The transmitted power through the air–antireflection (AR) coating–silicon interface is then evaluated using Snell's law and Fresnel coefficients. For blackbody illumination, the local transmittance is calculated as the average of the s- and p-polarized power transmittances. The resulting transmittance *T(u)*, shown in Fig. 4b, depends on the surface parameter $u$, which specifies the position on the ellipsoidal surface through the parametric relations $x(u)=a\sin(u)$ and $z(u)=b\cos u$.

Combining the projected-area weighting and the angle-dependent transmittance yields the optical power coupled into silicon as a function of surface position. The total transmitted power is obtained by integrating the transmitted power contribution over the illuminated ellipsoidal surface. Similarly, the incident power intercepted by the lens is calculated by integrating the projected incident power over the same surface. The overall transmission efficiency is then determined as the ratio between the total transmitted power and the total incident power. Using this approach, the calculated transmittance of the Si elliptical lens is 72.1% for the uncoated case and 92.6% after applying a Parylene-C AR coating.

The transmitted power from the Si elliptical lens is then mapped from the lens surface to the propagation angle inside silicon using the refracted ray direction obtained from Snell's law and the Fresnel's transmission coefficients of the air/Parylene-C/Si stack, while conserving the total transmitted power in the surface-to-angular coordinate transformation. The resulting scalar angular power density as a function of deflection angle $\theta$ is shown in Fig. 4c. Because of the high refractive index of silicon, the refracted angular power distribution is compressed and exhibits a cutoff at approximately $\theta\approx 73°$.

### *B. Elliptical Lens-Antenna Coupling Analysis*

For the coupling calculation, this scalar envelope was extended to spherical coordination by assuming azimuthal invariance of the elliptical lens intensity and was then converted into a transverse vector field by projecting the incident linear polarization onto the local $(\vec{e}_\theta, \vec{e}_\varphi)$ basis. The resulting $E_{\mathrm{ellip},\theta}$ and $E_{\mathrm{ellip},\phi}$ components were overlapped with the corresponding complex antenna fields on the same ($\theta$, $\phi$) grid. The final elliptical-lens–antenna coupling can therefore be calculated using the complex vector-basis overlap using eq. (2), including phase and polarization matching, with the unpolarized result obtained by the same averaging process reported in Section II.

The angular origin of the coupling is visualized in Fig. 5. The black and orange curves show the peak-normalized angular power contributions of the spiral antenna and the Si elliptical lens, respectively, after azimuthal integration. Their strong overlap indicates that the elliptical lens delivers most of its transmitted power into the angular region where the spiral antenna has a strong receiving response. This angular power matching is represented by the local scalar overlap envelope shown by the dashed blue curve in Fig. 5. The scalar coupling gives an upper-bound efficiency of $\eta_{power} = 90.1\%$, confirming that the angular power distributions of the elliptical lens and the antenna are very well matched when phase and polarization are assumed to be fully matched.

The coherent vectorial contribution was then evaluated using the same local complex inner product defined in Eq. (11), replacing the metalens field components with the corresponding elliptical-lens field components by integrating over $\phi$ at each fixed $\theta$, following the same procedure used for the metalens case. After global phase alignment, the resulting $\theta$-resolved contribution is plotted in Fig. 5. The green curve shows the contribution for the optimum incident linear polarization, while the pink curve shows the corresponding unpolarized contribution obtained by averaging orthogonal incident-polarization pairs following Eq. (3).

Using the complex vector-basis overlap in Eq. (2), the optimum linear-polarization coupling efficiency $\eta_{\text{vector,linpol}}$ is 43.4% , and the unpolarized coupling efficiency $\eta_{\text{vector,unpol}}$ following Eq. (3) is 42.0% . The apparent reduction of the raw vectorial coupling relative to the scalar upper bound mainly arises from the polarization-selective receiving mode of the spiral antenna, rather than from a large phase error introduced by the elliptical lens. After applying the per-receiving-mode correction following Eq. (4), the vectorial coupling remains close to the scalar phase-matched upper bound, with $\eta_{\text{unpol,coupling}} = 2\eta_{\text{vector,unpol}} = 84.0\%$, close to $\eta_{\text{scalar}} = 90.1\%$. This small remaining difference indicates that residual phase and polarization-vector mismatch between the elliptical-lens field and the spiral-antenna receiving mode is limited. This behavior is in clear contrast to the metalens case, where the high-deflection-angle region introduces stronger phase and polarization mismatch, leading to a much larger reduction of the vectorial coupling relative to the scalar or power-overlap limit. Therefore, the Si elliptical lens provides near-ideal quasi-optical coupling to the spiral antenna and serves as a useful reference for identifying the additional vectorial mismatch introduced by the metalens.

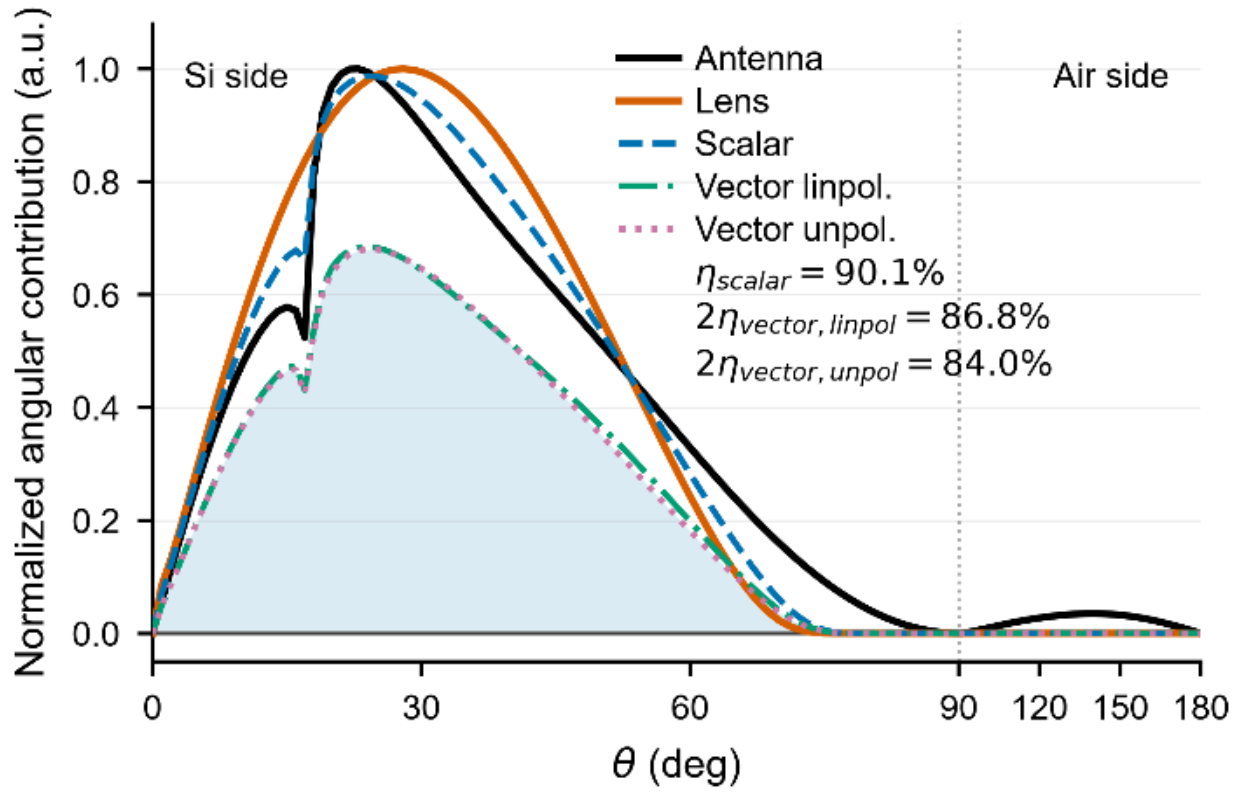

**Fig. 5.** Vectorial angular coupling analysis between the Si elliptical lens and spiral antenna. The black and orange curves show the peak-normalized angular power contributions of the simulated spiral antenna and the Si elliptical lens, respectively. The dashed blue curve represents the local scalar phase-matched overlap envelope, while the green and pink curves show the local vectorial overlap contributions for the optimum linear polarization and for unpolarized illumination, respectively.

## IV. Correlation Between Receiver Noise Temperature and Lens–Antenna Coupling Efficiency

The coupling-efficiency analysis developed in the previous sections can be directly validated by comparing it with receiver noise-temperature measurements. In our earlier experiments, the same HEB mixer, cryostat, optical setup, and local-oscillator configuration were used for both measurements, with the only significant change being the replacement of the dielectric coupling element from a conventional silicon elliptical lens to a silicon metalens. Therefore, the difference in measured receiver noise temperature can be primarily attributed to the different optical efficiencies of the two lens systems.

For the metalens configuration, the overall optical efficiency can be expressed as $\eta_{metalens} = \eta_{metalens-ant} \times \eta_{transmission}$, where $\eta_{metalens-ant}$ is the coupling efficiency between the spiral antenna and the metalens and $\eta_{transmission}$ is the transmission efficiency of the metalens. From the coupling analysis in Section II, the metalens-antenna coupling efficiency $\eta_{metalens-ant}$ is calculated to be 25.9 %. The transmission efficiency of the metalens $\eta_{transmission}$ can be estimated from the averaged transmittance of the meta atoms reported in ref. [18], yielding approximately 85%. Combining these factors gives a theoretical overall metalens efficiency of $\eta_{metalens} \approx$ 25.9% × 85% ≈22%.

For the conventional silicon elliptical lens, the efficiency can be expressed as $\eta_{ellip} = \eta_{ellip-ant} \times \eta_{transmission}$. Using the antenna–lens coupling efficiency calculated in Section III (84.0%) and the Fresnel-limited transmission efficiency of 92.6%, the theoretical efficiency of the elliptical lens is estimated to be $\eta_{ellip} \approx 84.0\% \times 92.6\% \approx 77.8\%$. Therefore, this theoretical analysis predicts that the metalens should achieve approximately 28.3% of the optical efficiency of the elliptical-lens system.

This prediction can be directly compared with the receiver-noise-temperature measurements. In the previous experiment, the receiver configuration was kept unchanged except for replacing the conventional silicon elliptical lens with the silicon metalens. The measured $T_{\text{rec}}^{\text{DSB}}$ were 530 K for the elliptical-lens-coupled receiver and 1800 K for the metalens-coupled receiver [18]. These values indicate that the effective optical efficiency of the metalens-based system is approximately 27% of that of the elliptical-lens system, in close agreement with the theoretical estimate of 28.3% obtained above. This agreement indicates that the reduced performance of the metalens receiver can be largely explained by the combined effects of reduced metalens–antenna vectorial coupling and finite metalens transmittance. It also supports the validity of the spherical-coordinate vectorial coupling analysis for quantitatively linking the optical design to the measured receiver performance.

## V. Perspective in Using Metalens for Heterodyne Receiver Array Applications

The analysis presented in Sections II–IV shows that the performance of the present metalens-coupled receiver is mainly limited by the reduced vectorial coupling and focusing efficiency at large deflection angles. Compared with a conventional silicon elliptical lens, the metalens must not only transmit the incident THz field but also impose a rapidly varying phase profile across the aperture. At large deflection angles, this requirement becomes increasingly demanding because the local phase gradient becomes steeper and the target wavefront must be sampled with higher spatial accuracy [24]. In principle, this requirement can be partially relaxed by employing a larger metalens diameter together with a thicker substrate spacer, thereby geometrically enlarging the effective phase sampling resolution. On the other hand, the implementation of rapidly varying phase profiles using discrete subwavelength meta-atoms introduces an intrinsic tradeoff between phase sampling fidelity and electromagnetic isolation between neighboring structures [25].

Reducing the meta-atom spacing improves the sampling of the target phase profile and suppresses parasitic diffraction, but densely packed meta-atoms also exhibit stronger near-field coupling. As a result, the optical response of each meta-atom can no longer be treated as locally independent and may deviate from the response predicted by the periodic-boundary-condition design library [18]. This effect is particularly severe in the outer high-deflection-angle region, where adjacent meta-atoms often have substantially different geometries. Consequently, the implemented wavefront deviates from the ideal design, causing scattering into non-focusing diffraction channels and producing additional phase and polarization mismatch in the coherent overlap with the spiral-antenna receiving mode.

Despite these challenges, recent advances in inverse-design methodologies have demonstrated promising routes toward overcoming the efficiency limitations of conventional metalens designs. Topology-optimized and freeform metalenses have achieved numerical apertures approaching 0.99 while maintaining substantially improved focusing efficiencies through nonlocal design strategies, multilayer architectures, and full-wave optimization techniques [16], [26]. These approaches directly address the phase-sampling and near-field-coupling limitations that dominate the performance of high-NA metalenses.

For heterodyne receiver systems, the relevant figure of merit is the observing efficiency, which approximately scales with $N \cdot \eta^2$ , where $N$ is the number of pixels and $\eta$ is the overall lens efficiency, considering the same receiver system with only different dielectric lenses [2]. This scaling implies that an increase in pixel count can compensate for a reduction in per-pixel optical efficiency. For example, a two-pixel metalens receiver array would achieve the same mapping speed as a single-pixel elliptical-lens receiver when the optical efficiency of each metalens pixel exceeds 55%. This threshold is obtained by equating the mapping speeds of the two systems and using the 77.8% optical efficiency calculated for the elliptical-lens receiver in Section IV.

The present metalens achieves a theoretical overall efficiency of 22%, which remains well below the 55% threshold required for a two-pixel metalens array to match the observing efficiency of a single-pixel elliptical-lens receiver. Based on the scaling relation $N \cdot \eta^2$, the current design would require approximately 13 pixels to surpass the performance of a single elliptical-lens receiver with a 77.8% optical efficiency. Nevertheless, the rapid progress of inverse-designed multilayer metasurfaces suggests that substantially higher efficiencies may be achievable in future generations of devices. Unlike conventional elliptical lenses, metalenses can be fabricated directly as dense planar arrays using standard nanofabrication technologies, enabling compact multi-pixel architectures that would be difficult to realize using conventional quasi-optical components. Consequently, while the present device does not yet provide an observing-efficiency advantage in low-pixel-count implementations, continued improvements in high-NA metasurface efficiency, combined with the inherent scalability of planar fabrication, may ultimately enable metalens-based heterodyne receiver arrays to outperform traditional single-pixel elliptical-lens systems.

## VI. Conclusion

In this work, we developed a quantitative spherical-coordinate framework for evaluating quasi-optical coupling between dielectric lenses and spiral antennas in THz HEB receivers. By representing the lens-delivered field and the antenna receiving mode in a common angular basis, the coupling efficiency can be evaluated from their complex vector overlap, including angular distribution, phase-front matching, and polarization projection. This approach provides a more rigorous metric than a scalar power-overlap estimate for determining how efficiently the optical mode delivered by the lens is coupled into the spiral-antenna/HEB receiving mode.

Applying this framework to a conventional silicon elliptical lens and a planar silicon metalens reveals the origin of their different receiver performances. For the elliptical lens, the scalar phase-matched upper bound is $\eta_{\mathrm{scalar}} = 90.1\%$, while the vectorial unpolarized coupling gives $\eta_{\mathrm{ellip-ant}} = 2\eta_{\mathrm{vector,unpol}} = 84.0\%$ . Combined with the calculated transmission efficiency of $92.6\%$ , this yields an overall elliptical-lens efficiency of approximately 77.8%. In contrast, the metalens–antenna vectorial coupling is calculated to be approximately $25.9\%$. Considering the averaged meta-atom transmission efficiency of $85\%$ , the theoretical overall metalens efficiency is approximately $22\%$ . Therefore, the present metalens system is expected to provide about $22\%/77.8\% \approx 28.3\%$ of the effective optical coupling efficiency of the elliptical-lens system. This theoretical ratio agrees well with the experimentally extracted ratio of 27% from receiver-noise-temperature measurements, where replacing the elliptical lens with the metalens increased $T_{\mathrm{rec}}^{\mathrm{DSB}}$ from 530 K to 1800 K [18]. The agreement shows that the measured receiver performance can be quantitatively explained by the reduced vectorial metalens–antenna coupling, together with metalens transmission efficiency.

Beyond explaining the measured receiver noise performance, the present analysis establishes a general design framework for evaluating and optimizing quasi-optical coupling in metalens-integrated heterodyne receivers. Although the current single-layer metalens does not yet reach the efficiency of a conventional elliptical lens, the analysis identifies the dominant limitations: high-deflection-angle focusing degradation, phase error, and polarization-vector mismatch in the coherent overlap with the antenna receiving mode. Recent advances in inverse-designed, multilayer, and nonlocal high-NA metalens provide realistic routes to improve these limitations. Combined with the intrinsic scalability of planar nanofabrication, such improvements could enable future metalens-based heterodyne receiver arrays to achieve mapping-speed advantages over traditional single-pixel elliptical-lens architectures.

## Acknowledgment

D. Ren would like to thank Dr. Jian-Rong Gao, Dr. Aurèle J.L. Adam, Dr. Alexei Semenov, Dr. Lei Liu, and Mr. Luuk Zonneveld for their helpful discussions and valuable insights.